\documentclass[
  journal=custom,
  manuscript=Research-Article, 
  year=2026,
  volume=00,
]{scntfcjrnl-journal}

\usepackage{microtype,siunitx,booktabs}
\title[A multilayer modelling framework for EHDV transmission in France]{A multilayer network-based modelling framework for epizootic haemorrhagic disease virus transmission in France}

\author{Viney Kumar}
\affiliation{Oniris, INRAE, BIOEPAR, Nantes, 44300, France}
\email[Viney Kumar]{viney.kumar@inrae.fr; vineykumar2801@gmail.com}

\author{Ga\"el Beaun\'ee}
\affiliation{Oniris, INRAE, BIOEPAR, Nantes, 44300, France}

\author{Sandie Arnoux}
\affiliation{Oniris, INRAE, BIOEPAR, Nantes, 44300, France}

\author{Fabien Corbi\`ere}
\affiliation{Univ Toulouse, ENVT, INRAE IHAP, Toulouse, France}

\author{Beno\^it Durand}
\affiliation{EPIMIM, Laboratoire de Santé Animale,
ANSES, Ecole Nationale Vétérinaire d'Alfort,
Maisons-Alfort, France}

\author{H\'el\`ene Cecilia}
\affiliation{Oniris, INRAE, BIOEPAR, Nantes, 44300, France}

\usepackage{tabularx}
\usepackage{tabularx}
\usepackage{array}
\usepackage{amsmath}
\usepackage{xurl}
\usepackage{float}
\setcitestyle{numbers,square,sort&compress}
\usepackage[colorlinks=true,
            citecolor=blue,
            linkcolor=black,
            urlcolor=blue]{hyperref}

\begin{document}

\renewcommand{\doi}[1]{\texttt{doi:\detokenize{#1}}}

\begin{abstract}
Epizootic haemorrhagic disease virus serotype 8 (EHDV-8) emerged in France in September 2023 and spread rapidly, but its spatial and temporal transmission dynamics remain poorly characterised. We adapted a stochastic, spatially explicit compartmental model originally developed for bluetongue to reconstruct the 2023--2024 EHDV-8 epidemic in cattle in mainland France. The model represented transmission within and between cantons, incorporating cattle demography, pasture cover, relative \textit{Culicoides} abundance and temperature-dependent vector processes. Three contact networks represented between-canton spread through short-range vector dispersal (pasture network), non-commercial cattle movements between pastures belonging to the same farm (farm network), and cattle trade (trade network). Using approximate Bayesian computation with sequential Monte Carlo (ABC-SMC), we calibrated the model to canton-week seroprevalence estimates derived from bulk-tank milk measurements. Calibration narrowed the plausible range of the baseline vector-mediated transmission parameter and favoured higher movement intensities through the farm network, although movement intensity remained uncertain. Simulation-based parameter recovery showed that calibration generally identified parameter regions capable of reproducing epidemic patterns comparable to the pseudo-observed data, although predictive performance varied among datasets and was typically lower than that obtained with the generating parameters. When applied to observed data, the calibrated model captured several features of the observed weekly median seroprevalence across sampled cantons, although discrepancies remained at some sampling times and predictive uncertainty was substantial. These results demonstrate how bulk-tank milk surveillance can inform the reconstruction of EHDV-8 epidemic patterns within a national-scale transmission framework. This framework provides a basis for future evaluation and optimisation of vaccination strategies.
\end{abstract}

\section{INTRODUCTION }
\label{sec:int}

Epizootic haemorrhagic disease (EHD) is a non-zoonotic vector-borne disease of domestic and wild ruminants caused by the epizootic haemorrhagic disease virus (EHDV), an arthropod-borne virus of the genus \textit{Orbivirus} \cite{jimenez2023epizootic,barua2024global}. The virus is primarily transmitted by biting midges of the genus \textit{Culicoides} and infects a wide range of ruminant species. EHD is often severe in wild deer, particularly white-tailed deer, whereas clinical disease is generally milder in cattle \cite{maclachlan2015epizootic}. Historically, the disease was restricted to the Americas, Africa, Asia and Australia \cite{jimenez2023epizootic,arbi2025emergence}. This situation changed in 2022 when EHDV serotype 8 (EHDV-8) was detected in Europe for the first time. Initial outbreaks were reported in Sicily and Sardinia, Italy, in October 2022 \cite{lorusso2023epizootic} and in southern Spain one month later \cite{ruiz2024emergence}. The virus subsequently spread across the Iberian Peninsula and reached France in 2023 \cite{zientara2024emergence,DEFRA2023EHD}.

In September 2023, the first reported cases of EHDV-8 were confirmed in cattle herds in the French administrative departments of Pyrénées-Atlantiques and Hautes-Pyrénées, near the Spanish border \cite{gondard2024exceptional}. The strain was closely related to viruses previously detected in Tunisia, Italy and Spain, supporting a progressive westward spread from North Africa into southern Europe \cite{gondard2024exceptional}. Following its introduction, EHDV-8 spread rapidly across southwestern France. By November 2023, more than 3,500 reported cases had been reported, and over 4,000 reported cases were recorded by the end of 2023 across 20 departments \cite{DEFRA2023EHD}. In response, movement restrictions and regulated zones were established to limit animal movements and reduce further dissemination of the virus \cite{MASAF2026EHD}. EHDV transmission declined markedly during the winter of 2023--2024, likely reflecting reduced \textit{Culicoides} abundance and activity under cooler temperatures \cite{hammami6461683climate}. New reported cases detected during the 2024 vector season indicated continued circulation of EHDV-8 and raised concerns regarding its ability to persist between transmission seasons \cite{DEFRA2024EHD}. To mitigate further spread, France launched a state-funded voluntary vaccination campaign in September 2024. The strategy relied on a 50-km-wide vaccination buffer zone along the internal boundary of the regulated zone, with the aim of achieving sufficient herd immunity to slow the progression of EHDV-8 \cite{MASAF2024Vaccination}. A total of 3,438 reported cases had already been reported in France by 12 December 2024 during the 2024--2025 reporting period \cite{DEFRA2024EHD}. The total subsequently reached 3,906 reported cases by 1 June 2025 \cite{MASAF2026EHD}. Only 5 reported cases were reported during the following period, between 1 June 2025 and 1 June 2026, and no additional reported cases have been reported since 1 June 2026 \cite{MASAF2026EHD}. Nevertheless, the factors underlying this decline, including the potential contribution of control measures, remain uncertain. A better understanding of EHDV-8 transmission dynamics is therefore important for informing future prevention and control strategies.

To date, EHDV transmission dynamics have received limited quantitative attention. Previous modelling studies have primarily focused on specific components of EHDV transmission. Gubbins et al. \cite{gubbins2024using} investigated the temperature-dependent transmission potential of EHDV in cattle and deer using the basic reproduction number, \(R_0\), while Fairbanks et al. \cite{fairbanks2024modelling} examined the influence of climatic conditions on EHDV vectorial capacity. Complementary studies have investigated the risk of EHDV-8 introduction and spread associated with long-distance dispersal. Bibard et al. \cite{bibard2024assessing} assessed the potential for windborne dispersal of infected \textit{Culicoides} during the emergence of EHDV-8 in France. More recently, Bibard et al. \cite{bibard2025spatial} also compared the risk of EHDV-8 introduction across Europe through windborne dispersal of \textit{Culicoides} and transboundary trade in live ruminants. However, the transmission dynamics underlying the recent spread of EHDV-8 within France remain poorly characterised. In particular, previous studies have not explicitly represented commercial animal movements, short-range non-commercial host movements, and vector-mediated transmission as distinct pathways within a dynamic transmission framework. Distinguishing these pathways is important for understanding their respective roles in shaping the spatial and temporal spread of EHDV-8.

In this study, we aimed to characterise the spatial and temporal dynamics of EHDV-8 transmission among cattle in mainland France. To this end, we adapted a stochastic, spatially explicit compartmental model originally developed for BTV spread in France \cite{courtejoie2019network}. The shared transmission of these orbiviruses by \textit{Culicoides} biting midges provided a biological basis for this adaptation \cite{anthonioz2025heterogenous,hammami6461683climate}. The model represents infection dynamics within and between cantons, French administrative units below the department level, at weekly time steps. Three distinct network layers represent between-canton spread through commercial cattle movements (hereafter, the trade network), non-commercial movements between pastures belonging to the same farm (hereafter, the farm network), and short-range vector dispersal (hereafter, the pasture network). Temperature and pasture cover influence vector-mediated transmission, allowing the model to account for seasonal variation and spatial heterogeneity in infection pressure.

The initial model represented BTV spread at the canton level during the 2007--2009 outbreak and through 2010. However, the French canton system was subsequently redrawn, with several canton boundaries modified following the 2015 territorial reform \cite{bourdin2021territorial}. Applying this framework to EHDV-8 required updating both its disease-specific parameters and its underlying data layers. We reconstructed the spatial framework using current canton boundaries and incorporated contemporary cattle population and birth records to represent host demography. Pasture locations and their association with farms defined the pasture and farm networks, while recorded cattle movements informed the trade network. Updated temperature and pasture-cover data characterised the environmental conditions influencing transmission. We then used paired individual serological results and bulk-tank milk competitive enzyme-linked immunosorbent assay (ELISA) measurements to model the relationship between bulk-tank milk inhibition values and herd-level seroprevalence using beta-binomial regression \cite{mamun2023model}. Predicted herd-level seroprevalence was aggregated by canton and week and used to calibrate the transmission model through approximate Bayesian computation with sequential Monte Carlo (ABC-SMC) \cite{lenormand2013adaptive}. Inference focused on two parameters: the baseline vector-mediated transmission parameter ($\beta_0$) and the farm-network movement scaling parameter ($\Psi_F$). Both parameters were estimated because reliable EHDV-specific estimates were unavailable in the literature. We assessed their practical identifiability through simulation-based parameter recovery and used posterior predictive checks to evaluate agreement with the seroprevalence estimates. Together, these analyses provide a data-informed framework for reconstructing the spatial and temporal dynamics of the EHDV-8 epidemic in mainland France while accounting for uncertainty in key transmission parameters.

\begin{table*}[t]
\centering
\caption{Description of the parameters used in the model for the epizootic haemorrhagic disease.}
\label{tab:fixed_parameters}

\renewcommand{\arraystretch}{1.35}

\begin{tabularx}{\textwidth}{|
>{\centering\arraybackslash}m{1.3cm}|
>{\raggedright\arraybackslash}X|
>{\raggedright\arraybackslash}m{6cm}|
>{\raggedright\arraybackslash}m{3cm}|
}
\hline
\textbf{Symbol} & \textbf{Description} & \textbf{Value} & \textbf{Reference} \\
\hline

$\lambda_1$ &
1/ length of persistence of colostral antibodies (rate) &
0.0476 (1/21 wk) & \cite{toye2013bluetongue}
\\ \hline

$\lambda_2$ &
Literature-derived recovery rate corresponding to the mean viraemic duration &
$1.031~\mathrm{week}^{-1}$ (mean duration: $0.97$ week) &
\cite{gubbins2024using}
\\ \hline

$b(c,t)$ &
biting rate $(week^{-1})$ &
$
\begin{array}{l}
7 \cdot a_0 \cdot T(c,t) \cdot (T(c,t)-T_0)\cdot (42-T)^{1/2}
\\[0.2cm]
\text{with:}
\\[0.1cm]
a_0 = 1.4 \times 10^{-4}
\quad (9.8\times10^{-5},\,2.2\times10^{-4})
\\[0.15cm]
T_0 = 6.93
\quad (0.49,\,14.32)
\end{array}
$
&
depends on temperature; parameters ($a_0$ and $T_0$ estimated in \cite{gubbins2024using})
\\ \hline

$\mu^v(c,t)$ &
daily mortality proportion of \textit{Culicoides} vectors &
$
\begin{array}{l}
\mu_0 \cdot exp(\mu_1 \cdot T(c,t))
\\[0.2cm]
\text{with:}
\\[0.1cm]
\mu_0 = 0.012
\quad (0.007,\,0.019)
\\[0.15cm]
\mu_1 = 0.15
\quad (0.13,\,0.17)
\end{array}
$
& depends on temperature; parameters ($\mu_0$ and $\mu_1$ estimated in \cite{gubbins2024using})
\\ \hline

$T_{min}$ &
threshold temperature for virus replication &
14.97$^\circ$C & threshold temperature estimated in \cite{gubbins2024using}
\\ \hline

$\Psi_P$ &
proportion of canton surface reachable in a week by vectors from neighbouring cantons &
0.290 & Flight distance \cite{kluiters2015local,courtejoie2019network}, canton surface (Supplement S4)
\\ \hline

$N_{inf}$ &
number of infected cattle introduced to seed infection &
5 & \cite{courtejoie2019network}
\\ \hline
$z_{\max}$ & Maximum mean canton elevation for vector-mediated transmission
& 500~m & \cite{courtejoie2019network} \\
\hline
\end{tabularx}
\end{table*}

\section{Material and methods}
\subsection{Study area and study period}
We studied the spread of EHDV-8 in mainland France, excluding Corsica and overseas territories, from mid-September 2023 (model week 36) to 29 December 2024--4 January 2025 (model week 104). This period encompassed the 2023–2024 EHDV-8 epidemic and extended through the end of the voluntary vaccination campaign. Cantons were used as the spatial units of the model. During the study period, mainland France comprised 1,967 cantons, each containing an average of 18 municipalities and covering approximately 272 km$^2$.
\subsection{Data collection}
The cattle population data and all cattle trade movements between pairs of cantons from 2023 to 2025 were extracted from the French National Identification Database (BDNI), an exhaustive database maintained by the Ministry of Agriculture. Pasture data were derived from the 2024 French Registre Parcellaire Graphique (RPG), the official parcel-level agricultural land-use database maintained under the European Union Common Agricultural Policy (CAP) framework (see Supplementary S1.1). The publicly available RPG data were used to characterise pasture area, with pastures defined as grasslands, either permanent or temporary when included in a grass–arable rotation system. Additional information on pasture locations and the identification of pastures belonging to the same farm was obtained upon request from the RPG Level 2 dataset provided by the ODR Service Unit, INRAE Occitanie–Toulouse. Temperature data were obtained from the SAFRAN atmospheric analysis system maintained by Météo-France, with a spatial resolution of $0.072^\circ$ (8 km). Daily temperature records covering the period 2023–2025 were extracted for subsequent analyses (see Supplementary S1.1). 
Elevation data were obtained from the Copernicus Digital Elevation Model (COP-DEM) through the OpenTopography data portal, with an approximate spatial resolution of 30 m. Mean elevation was calculated for each canton by extracting raster values within official canton boundaries and computing their arithmetic mean. All spatial datasets were aggregated at the canton level, whereas temporal data were aggregated on a weekly basis.

Serological data were provided by researchers at ENVT and INRAE, Toulouse, and comprised repeated bulk-tank milk competitive ELISA measurements and individual blood or milk ELISA results from a subset of herds. The wider surveillance dataset covered more than 1,100 dairy herds across 15 French departments, mainly in Occitanie, Normandy, Pays de la Loire and Grand Est, from late 2023 to early 2025. Sampling schedules varied between regions. Further details of the data sources, sampling coverage and preparation of the paired and control datasets are provided in Supplementary Section~S5.1. 

\subsection{Model design}
\label{sec:model_design}

\subsubsection{Spatial and temporal modelling framework}
\label{subsec:spatiotemporal_framework}

\begin{figure*}[t]
    \centering
    \includegraphics[
    width=0.95\textwidth,
    keepaspectratio
    ]{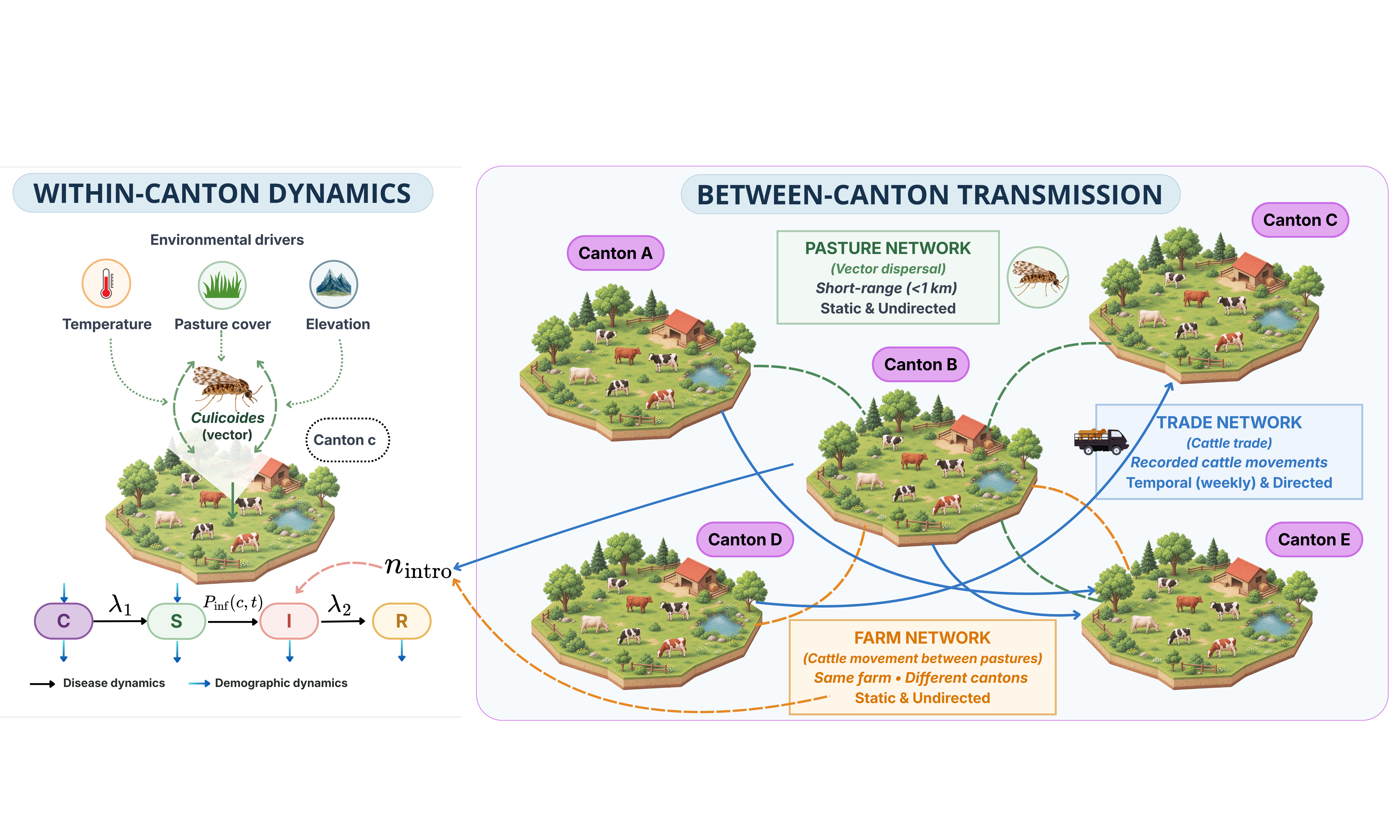}

    \caption{Schematic representation of the EHDV-8 transmission model. The cattle population of each canton $c$ is partitioned into four epidemiological compartments:
    $C$, cattle protected by maternally derived colostral antibodies;
    $S$, susceptible cattle;
    $I$, infectious cattle; and
    $R$, recovered cattle with persistent anti-EHDV antibodies. Spatial spread between cantons is represented by three network layers. The farm and trade networks generate movement-mediated infectious introductions into destination cantons; the total number of such introductions during week $t$ is denoted by $n_{\mathrm{intro}}(c,t)$.}

    \label{fig:schematic}
\end{figure*}

Our EHDV-8 modelling framework was adapted from the stochastic, 
spatially explicit compartmental model originally developed to describe 
bluetongue virus transmission in mainland France 
\cite{courtejoie2019network}. The model represents EHDV-8 transmission 
among cattle at the canton level using a discrete weekly 
time step. This temporal resolution was retained from the original modelling framework, 
which was structured around weekly demographic, movement, and transmission 
processes. However, the duration of viraemia in cattle is considerably shorter for EHDV-8 and has been estimated at approximately 0.97 weeks \cite{gubbins2024using}. Because this duration closely matches the one-week temporal resolution of the model, viraemia was represented as lasting approximately one model time step. Accordingly, cattle present in the infectious ($I$) compartment were parameterised to recover with an approximately unit probability at the subsequent weekly update. This discrete-time approximation allowed the reported EHDV-8 viraemic duration to be represented without reformulating the model at a finer temporal resolution.  Within each canton, the cattle population is partitioned into 
epidemiological compartments according to the infection status of the 
animals (Figure ~\ref{fig:schematic}).

The cattle population size in each canton was initialised using observed 
demographic data from the BDNI. Demographic changes were subsequently 
updated at each weekly time step using the observed number of cattle and 
the recorded number of births, rather than a modelled birth rate. To ensure 
consistency between these two quantities, demographic flows were reconciled 
with the observed cattle population at each time step. When the sum of the cattle population from the preceding week and the recorded births exceeded the current observed population, the difference was treated as animal exits. Conversely, when this sum was lower than the observed population, the number of births was adjusted accordingly. This reconciliation ensured that the modelled cattle population matched the observed population at each weekly time step.

Rather than explicitly representing the vector population using separate 
vector compartments, vector-mediated transmission was represented through 
the force of infection. The model retains the infectious cattle prevalence 
from previous weeks to account for the delayed contribution of vectors that 
became infected during those weeks. Their contribution to transmission in 
the current week depends on temperature-dependent vector survival and 
completion of the extrinsic incubation period. Thus, the current 
vector-mediated force of infection depends partly on infection occurring 
during previous weeks, as described in detail below (see \ref{subsec:within_canton_foi}).

\subsubsection{Initial conditions and epidemic seeding}
\label{subsec:model_initialisation}

At the beginning of the simulation, all cattle were assumed to be
susceptible. The epidemic was seeded during the week corresponding to
the first confirmed detection of EHDV-8 in France in mid-September 2023.
Infection was introduced into the seven cantons in which EHDV-8 was first
confirmed. As the actual number of infectious cattle at the time of
detection was unknown, five infectious cattle were introduced into each
canton as a pragmatic modelling assumption, following the initial seeding
approach used in the original BTV modelling framework
\cite{courtejoie2019network}. Thus, a total of 35 infectious cattle were
used to initiate the simulated epidemic.

No additional exogenous seeding was imposed at the beginning of subsequent 
transmission seasons. Instead, the infection dynamics were allowed to 
evolve continuously according to the host compartmental transitions, 
cattle movements, and temperature-dependent vector processes represented 
in the model.

\subsubsection{Weekly infection process}
\label{subsec:weekly_infection_process}

Let $c$ denote a canton and $t$ denote a weekly time step. During each 
week, the total number of cattle added to the infected population in 
canton $c$, denoted by $n_{\mathrm{inf}}(c,t)$, was decomposed into two 
components:

\begin{equation}
n_{\mathrm{inf}}(c,t)
=
n_{\mathrm{vec}}(c,t)
+
n_{\mathrm{intro}}(c,t),
\label{eq:total_infections}
\end{equation}

where $n_{\mathrm{vec}}(c,t)$ is the number of susceptible cattle newly 
infected through infectious vector bites, and 
$n_{\mathrm{intro}}(c,t)$ is the number of already infectious cattle 
introduced into canton $c$ during week $t$.

$n_{\mathrm{intro}}(c,t)$ accounts 
for infectious cattle entering the canton through movements between trade network or farm network.

The number of vector-mediated infections was sampled from a binomial 
distribution:

\begin{equation}
n_{\mathrm{vec}}(c,t)
\sim
\operatorname{Binomial}
\left(
S(c,t),
P_{\mathrm{inf}}(c,t)
\right),
\label{eq:vector_infections}
\end{equation}

where $S(c,t)$ is the number of susceptible cattle in canton $c$ at week 
$t$, and $P_{\mathrm{inf}}(c,t)$ is the probability that a susceptible 
animal becomes infected through vector bites during that week.

Assuming that the infection pressures generated by local vectors and by 
vectors associated with neighbouring cantons act additively, the weekly 
probability of infection was defined as

\begin{equation}
P_{\mathrm{inf}}(c,t)
=
1-
\exp
\left[
-
\left(
\lambda_{\mathrm{int}}(c,t)
+
\lambda_{\mathrm{ext}}(c,t)
\right)
\right],
\label{eq:infection_probability}
\end{equation}

where $\lambda_{\mathrm{int}}(c,t)$ is the force of vector-borne infection 
generated by female \textit{Culicoides} located within canton $c$, and 
$\lambda_{\mathrm{ext}}(c,t)$ is the force of infection generated by 
vectors associated with other cantons located within the assumed flight 
distance of canton $c$.

Thus, susceptible cattle may become infected either through infectious 
vectors originating within their own canton or through infectious vectors 
contributing infection pressure from surrounding cantons.

\subsubsection{Within-canton vector-borne force of infection}
\label{subsec:within_canton_foi}

The within-canton force of infection, 
$\lambda_{\mathrm{int}}(c,t)$, represents the infection pressure generated 
by female \textit{Culicoides} that acquired EHDV while feeding on infectious 
cattle in canton $c$ during previous weeks. For these vectors to contribute 
to transmission at week $t$, they must have completed the extrinsic 
incubation period required for viral replication and dissemination to the 
salivary glands and must have survived until week $t$.

Rather than modelling the absolute vector-to-host ratio explicitly,
spatial and temporal variation in vector availability was represented
through the canton--week relative-abundance index described below. Under this simplifying assumption, 
the within-canton vector-borne force of infection can be represented as a non-Markovian process:

\begin{equation}
\lambda_{\mathrm{int}}(c,t)
=
\tau(c,t) \times
\sum_{i=1}^{13}
w_i(c,t)\, \times
\mathrm{Prev}(c,t-i),
\label{eq:within_canton_foi}
\end{equation}

where $\mathrm{Prev}(c,t-i)$ is the proportion of infectious cattle in canton $c$ during week $t-i$. The model retains the modelled infectious cattle prevalence from the preceding 13 weeks (approximately three months). 

The weight $w_i(c,t)$ scales the contribution of infectious cattle
prevalence in week $t-i$ to the vector-borne force of infection
acting on susceptible cattle in week $t$. Susceptible \textit{Culicoides} may acquire EHDV when feeding on infectious cattle in week $t-i$. Midges infected at that time can subsequently transmit the virus to susceptible cattle in week $t$ if they have survived and completed the extrinsic incubation period. Each weight therefore combines the modelled EIP-completion fraction with the probability of vector survival over $i$ weeks (see Supplementary Section~S3). This delayed transmission process is represented without explicitly modelling vector infection compartments. 

The between-canton component, $\lambda_{\mathrm{ext}}(c,t)$, represents the vector-borne infection pressure on susceptible cattle in canton $c$ originating from neighbouring cantons within the assumed dispersal range of \textit{Culicoides}. This component accounts for short-range vector dispersal between cantons through the pasture network.

\subsubsection{Effective vector--host contact rate}
\label{subsec:effective_contact_rate}

The effective contact rate between cattle and vectors in canton $c$ during 
week $t$ was defined as

\begin{equation}
\tau(c,t) = \beta_0 \times \tilde{A}(c,t) \times Env(c) \times b(c,t),
\label{eq:effective_contact_rate}
\end{equation}

where $\beta_0$ is the baseline vector-mediated transmission parameter,
$Env(c)$ is a canton-specific environmental proxy, and $b(c,t)$ is the
temperature-dependent biting rate of \textit{Culicoides}.

The coefficient $\beta_0$ summarises the baseline vector-to-host ratio and
the probabilities of successful host-to-vector and vector-to-host
transmission, which were not parameterised separately. To introduce spatial
and temporal heterogeneity in vector availability, we used daily gridded
mean predictions of total \textit{Culicoides} abundance per
trap-night \cite{hammami6461683climate}. Predictions were averaged within
model weeks and spatially aggregated to cantons according to the surface
area of overlap between raster cells and canton boundaries; full details of
the spatial and temporal processing are provided in Supplementary
Section~S1.2. The resulting canton-week abundance, $A(c,t)$, was
normalised by the maximum value across all included cantons and model weeks:

\begin{equation}
\widetilde{A}(c,t)
=
\frac{A(c,t)}{A_{\max}},
\qquad
A_{\max}
=
\max_{c,t} A(c,t).
\end{equation}

The resulting dimensionless relative-abundance index preserved spatial and
temporal differences in predicted vector availability and was used as a
relative proxy rather than as a direct estimate of the absolute
vector-to-host ratio.

The environmental proxy, $\mathrm{Env}(c)$, was defined using the 
proportion of the canton surface covered by pasture. This formulation was 
based on the assumption that EHDV transmission is more likely in areas 
where pasture availability increases the spatial co-occurrence of cattle 
and \textit{Culicoides}. Therefore, pasture coverage was used as a proxy 
for host availability and for environmental conditions associated with 
\textit{Culicoides} presence and abundance. 

Vector-mediated transmission was assumed to be negligible in cantons
with a mean elevation of 500~m or higher and was therefore set to zero
at or above this threshold \cite{courtejoie2019network}.

Finally, $b(c,t)$ represents the biting activity of \textit{Culicoides} 
during week $t$ in canton $c$. Because the biting rate depends on 
temperature, this component introduces seasonal and geographical variation 
in vector activity. Together, $\beta_0$, $\mathrm{Env}(c)$, $b(c,t)$, and the 
temperature-dependent weights $w_i(c,t)$ allow the model to represent 
spatial heterogeneity in vector suitability, seasonal changes in biting 
activity, vector survival, and the delayed effect of the extrinsic 
incubation period on EHDV-8 transmission.

\subsubsection{Between-canton transmission through a multilayer contact network}
\label{subsec:between_canton_networks}

The spatial spread of EHDV-8 between cantons was represented using a 
multilayer contact network comprising three epidemiologically distinct 
layers: the pasture network, the farm network, and the trade network. 
In each network, nodes represented cantons; however, the biological 
interpretation, directionality, and temporal structure of the links differed 
among network layers. The pasture network represented the short-range 
dispersal of infectious \textit{Culicoides}, whereas the farm and trade 
networks represented cattle movements capable of introducing already 
infectious animals into destination cantons 
\cite{courtejoie2019network}. The pasture network was constructed to represent the local dispersal of 
\textit{Culicoides} between geographically adjacent pasture areas. An 
undirected link was defined between two cantons when at least one pair 
of pastures, with one pasture located in each canton, was separated by 
less than 1~km. This distance was selected following the criterion used 
by Palisson et al. \cite{palisson2017analysis} to represent the most likely pathways of vector-borne disease transmission across the densely connected network of French pastures. The farm network represented cattle movements between geographically separated pastures belonging to the same farm. Two cantons were connected when at least one farm managed pastures located in both cantons. Because 
cattle could be moved between the different pastures of the same farm in 
either direction, the farm network was represented as a static and 
undirected network. The trade network represented recorded cattle movements between farms. A directed link from a source canton $j$ to a destination canton $c$ was 
present during week $t$ when cattle were traded from at least one farm 
located in canton $j$ to a farm located in canton $c$. In contrast to the 
pasture and farm networks, the trade network was therefore both directed 
and time varying, with the identities of donor and recipient cantons, as 
well as the number of cattle moved, updated at every weekly time step. The pasture and farm networks remained static throughout the simulation, 
and their links could support transmission in either direction. By contrast, 
the trade network changed from week to week according to the observed 
cattle movement records. The topological properties of the three network 
layers are described in Supplementary S8.

\subsubsection{Vector dispersal through the pasture network}
\label{subsubsec:vector_dispersal}

The pasture network contributed to EHDV-8 transmission by allowing 
infection pressure generated by infectious vectors in one canton to affect 
susceptible cattle in neighbouring cantons. Let $\mathcal{N}_{P}(c)$ denote 
the set of cantons connected to the canton $c$ through the pasture network. 
The force of vector-borne infection contributed by neighbouring cantons was 
defined as

\begin{equation}
\lambda_{\mathrm{ext}}(c,t)
=
\Psi_{P} \times
\sum_{j \in \mathcal{N}_{P}(c)}
\lambda_{\mathrm{int}}(j,t),
\label{eq:between_canton_vector_foi}
\end{equation}

where $\lambda_{\mathrm{int}}(j,t)$ is the within-canton force of 
vector-borne infection generated in the source canton $j$ during week $t$, 
and $\Psi_{P}$ is the proportion of the destination canton assumed to be 
reachable within one week by vectors arriving from each neighbouring canton (see Supplementary S4). Thus, each neighbouring canton contributed a fraction 
$\Psi_{P}\lambda_{\mathrm{int}}(j,t)$ of its locally generated vector 
infection pressure to canton $c$.

\subsubsection{Host-mediated introductions through the farm and trade networks}
\label{subsubsec:host_movement_introductions}

Unlike transmission through the pasture network, the farm and trade
networks introduced cattle that were already infectious before entering
the destination canton. Let $j$ denote a source canton and $c$ a
destination canton. The weekly movement volumes associated with movements
from $j$ to $c$ through the farm and trade networks are denoted by
$N_{\mathrm{F}}(c,j,t)$ and $N_{\mathrm{T}}(c,j,t)$, respectively.

For movements between pastures belonging to the same farm, the modelled
weekly movement volume was defined as

\begin{equation}
N_{\mathrm{F}}(c,j,t)
=
\left\lfloor
\Psi_{F}
Q_{\mathrm{F}}(c,j)
N(j,t)
\right\rfloor ,
\label{eq:farm_movement_volume}
\end{equation}

where $\Psi_F$ is a farm-network movement scaling parameter,
$Q_F(c,j)$ is the proportion of cattle farms in source canton $j$
that also manage at least one pasture in destination canton $c$,
and $N(j,t)$ is the cattle population of source canton $j$ during
week $t$. The pair-specific fraction used to determine the modelled
movement volume from $j$ to $c$ is therefore $\Psi_F Q_F(c,j)$.
The resulting movement volume was converted to an integer number of
animals by truncation before the binomial sampling step.

For the trade network, $N_{\mathrm{T}}(c,j,t)$ was obtained directly
from the observed weekly cattle movement records extracted from the BDNI. The trade network therefore retained the observed direction and number of cattle movements between each source--destination pair. During the EHDV-8 epidemic, movement restrictions were implemented in France, including regulated zones extending 150 km around confirmed outbreaks and conditions for moving animals out of these zones. Unlike Courtejoie et al. (\cite{courtejoie2019network}), we did not introduce a separate parameter for compliance with movement restrictions, as the observed movement records already reflected the cattle movements that occurred under the prevailing regulatory conditions. Consequently, the model accounted for the observed trade patterns without explicitly representing the restrictions or estimating their compliance.

Assuming that animals associated with each movement were sampled randomly
with respect to infection status, the number of infectious introductions
generated through network $r$, where
$r \in \{\mathrm{F},\mathrm{T}\}$, was drawn from a binomial distribution:

\begin{equation}
n_{r}(c,j,t)
\sim
\operatorname{Binomial}
\left(
N_{r}(c,j,t),
\mathrm{Prev}(j,t)
\right),
\qquad
r \in \{\mathrm{F},\mathrm{T}\},
\label{eq:infectious_movers}
\end{equation}

where $\mathrm{Prev}(j,t)$ is the proportion of infectious cattle in
source canton $j$ during week $t$.

Let $\mathcal{N}_{\mathrm{F}}(c)$ denote the source cantons connected to
$c$ through the farm network, and let
$\mathcal{N}_{\mathrm{T}}^{-}(c,t)$ denote the set of donor cantons with
an incoming trade connection to $c$ during week $t$. The total number of movement-mediated infectious introductions into
canton $c$ was then

\begin{equation}
n_{\mathrm{intro}}(c,t)
=
\sum_{j \in \mathcal{N}_{\mathrm{F}}(c)}
n_{\mathrm{F}}(c,j,t)
+
\sum_{j \in \mathcal{N}_{\mathrm{T}}^{-}(c,t)}
n_{\mathrm{T}}(c,j,t).
\label{eq:movement_introductions}
\end{equation}

Movement-mediated infectious introductions were represented as
transitions from the susceptible to the infectious compartment in the
destination canton, such that the total cattle population of the
destination canton remained unchanged. Outgoing infectious animals were not explicitly removed from the source
canton, as their depletion was assumed to have a negligible effect on
within-canton infection dynamics.

Most model parameters were fixed using estimates obtained from the
published literature or values directly informed by empirical data
(Table~\ref{tab:fixed_parameters}). Reliable EHDV-specific values were
unavailable for two model parameters, which were therefore estimated
from the available EHDV-8 data using the inference procedure described
in Section~\ref{sec:parameter_estimation}.
\subsection{Relationship between bulk-tank milk inhibition and herd seroprevalence}

Bulk-tank milk analyses were available for a larger number of herds and dates than individual-level serological data but only provided inhibition values rather than direct estimates of herd seroprevalence. To reconstruct herd-level seroprevalence for these herds, we quantified the relationship between bulk-tank milk inhibition values as determined by a competitive ELISA and within-herd seroprevalence using paired data.

We used two datasets. The first comprised herds for which both individual-level serological analyses and bulk-tank milk measurements were available. For each bulk-tank milk analysis date, the corresponding herd seroprevalence was estimated from the closest individual sampling date, retaining only observations with the smallest absolute delay between the two dates. The second dataset comprised control herds from departments (76 and 55) with no evidence of EHD circulation and no vaccination, for which seroprevalence was assumed to be zero. Because multiple bulk-tank milk measurements were available for some control herds, inhibition values were averaged at the herd level so that each control herd contributed a single value to the analysis.

Let $Y_i$ denote the number of seropositive animals in herd $i$, $n_i$ the number of animals tested, $p_i$ the corresponding herd seroprevalence, and $x_i$ the bulk-tank milk inhibition value. We considered the following model structure:
\begin{equation}
Y_i \sim \mathrm{Binomial}(n_i, p_i)
\end{equation}
or
\begin{equation}
Y_i \sim \mathrm{Beta\text{-}binomial}(n_i, p_i, \phi),
\end{equation}
with
\begin{equation}
\mathrm{logit}(p_i) = \gamma_0 + \gamma_1 x_i.
\end{equation}

For control herds, individual serological data were not available, so the number of animals that would have contributed to the seroprevalence estimate was unknown. To incorporate these herds under the same response structure as herds with individual-level data, the numerator $Y_i$ was set to zero, and the denominator $n_i$ was approximated by the rounded mean number of lactating cattle contributing to the bulk-tank milk samples for each control herd across its available observations. This allowed control herds to contribute to the estimation of the inhibition--seroprevalence relationship while preserving the same response structure as in infected herds.

Both models were fitted by maximum likelihood. Alternative formulations, including the use of all control observations and constrained models enforcing near-zero seroprevalence at maximum inhibition, were also evaluated and are described in the Supplementary Material S5.3. Model selection was based on the corrected Akaike Information Criterion (AICc), with results summarised in Supplementary Tables~S1 and S2.

\subsection{ABC-SMC for parameter estimation}
\label{sec:parameter_estimation}

In this modelling framework, we adopted the Approximate Bayesian Computation with Sequential
Monte Carlo sampling (ABC-SMC) method for parameter estimation \cite{toni2009approximate,delmoral2012adaptive}.
We estimated two parameters:

\begin{equation}
\boldsymbol{\theta}
=
\left(
\beta_{0},
\Psi_{F}
\right),
\label{eq:estimated_parameter_vector}
\end{equation}

where $\beta_{0}$ represents the baseline vector-mediated transmission parameter and $\Psi_{F}$ is the farm-network movement scaling
parameter. All remaining epidemiological and transmission parameters were fixed at the EHDV-8 values reported in Table~\ref{tab:fixed_parameters}. Independent uniform prior distributions were assigned to the two
parameters:

\begin{equation}
\beta_{0} \sim \mathcal{U}(10,30000),
\qquad
\Psi_{F} \sim \mathcal{U}(0,1).
\label{eq:parameter_priors}
\end{equation}

The epidemic simulation was initialised at model week 36 (10--16 September 2023), corresponding to mid-September 2023, and continued through week 104 (29 December 2024--4 January 2025). Parameter estimation was based on observations from weeks 76--104 (16 June 2024--4 January 2025) because the bulk-tank milk seroprevalence data were available primarily from June 2024 onwards. The calibration target consisted of the complete set of available canton-week seroprevalence values; no additional dimension-reducing summary statistics were calculated.

An observation model was used to compare model outputs and observation-derived seroprevalence on the same scale. For a proposed parameter vector $\boldsymbol{\theta}$, one stochastic epidemic realisation was generated. Cattle in the recovered compartment were assumed to represent the seropositive 
population, with seroconversion approximated at recovery following the 
approximately one-week viraemic period and antibodies assumed to persist 
thereafter. Infectious cattle and calves with maternally derived colostral antibodies were not considered seropositive. Calves with colostral antibodies were excluded because the serological data were derived from bulk-tank milk
samples from lactating cows, in which colostral
antibodies would no longer be present. In canton $c$ and week $w$, the latent model-based seroprevalence was calculated as the proportion of cattle in the recovered compartment

\begin{equation}
p^{\mathrm{lat}}_{cw}(\boldsymbol{\theta})
=
\frac{
R_{cw}(\boldsymbol{\theta})
}{
N_{cw}(\boldsymbol{\theta})
},
\label{eq:main_latent_seroprevalence}
\end{equation}

where $R_{cw}(\boldsymbol{\theta})$ is the number of recovered cattle
and $N_{cw}(\boldsymbol{\theta})$ the total cattle population. This canton-level seroprevalence was used as the underlying
seropositivity probability for the corresponding serological observations. To reproduce the sampling variability in the reference data, a simulated number of seropositive cattle was drawn using the canton-week sampled population $n_{cw}$:

\begin{equation}
Z^{\mathrm{sim}}_{cw}(\boldsymbol{\theta})
\sim
\operatorname{Binomial}
\left[n_{cw}, p^{\mathrm{lat}}_{cw}(\boldsymbol{\theta}) 
\right],
\qquad
\widehat{p}^{\mathrm{sim}}_{cw}(\boldsymbol{\theta})
=
\frac{
Z^{\mathrm{sim}}_{cw}(\boldsymbol{\theta})
}{
n_{cw}
}.
\label{eq:main_simulated_observation}
\end{equation}

Diagnostic sensitivity and specificity were not explicitly incorporated into the observation model, as the transmission model simulated cattle infection states rather than bulk-tank milk inhibition values or ELISA test outcomes. The discrepancy associated with $\boldsymbol{\theta}$ was the sum of
squared differences between the simulated sampled seroprevalence and
the reference seroprevalence over all observed canton-week pairs:

\begin{equation}
\rho(\boldsymbol{\theta})
=
\sum_{(c,w)\in\mathcal{O}}
\left[
\widehat{p}^{\mathrm{sim}}_{cw}(\boldsymbol{\theta})
-
p^{\mathrm{obs}}_{cw}
\right]^{2},
\label{eq:main_abc_distance}
\end{equation}

where $\mathcal{O}$ denotes the set of canton-week pairs retained for
inference and $p^{\mathrm{obs}}_{cw}$ is the corresponding reference
seroprevalence.

The ABC-SMC procedure was implemented with the
\texttt{BRREWABC} R package
\cite{beaunee2024brrewabc}.

Each complete generation comprised 5,000 accepted particles. After the
initial generation, the tolerance for the next generation was set to
the $0.8$ quantile of the particle distances obtained in the preceding
generation. The default weighted-resampling and adaptive Gaussian
perturbation procedures implemented in \texttt{BRREWABC} were used.
One stochastic model realisation was evaluated per proposed particle.

A maximum of 30 generations was requested. The procedure was stopped
during generation 23 because the target of 5,000 accepted particles
could not be reached before the maximum limit of 100,000 proposed
particles for that generation was attained. Consequently, posterior inference was based on the weighted particle
population from the last fully completed generation. 

We assessed the ability of the ABC-SMC procedure to recover model parameters and reproduce simulated epidemic patterns using 50 pseudo-observed datasets generated under known parameter combinations selected by Latin hypercube sampling (LHS) \cite{song2023monte}. For each combination, 100 stochastic epidemic realisations were generated, and a representative realisation was selected based on its agreement with the pointwise median trajectory of weekly seroprevalence across sampled cantons. Each pseudo-observed dataset was independently calibrated using ABC-SMC, and predictive performance was assessed by comparing simulations generated from the resulting posterior distribution with those generated using the known parameter values. Details of the ABC-SMC implementation and the simulation-based assessment are provided in Supplementary Sections~S6 and~S7, respectively.

For clarity, the main steps linking the serological data to parameter estimation for the observed EHDV-8 epidemic are summarised below:

\begin{enumerate}
    \item \textbf{Construction of reference seroprevalence.}
    A beta-binomial regression model was fitted to paired individual serological results and bulk-tank milk inhibition values measured by competitive ELISA to establish the relationship between inhibition and within-herd seroprevalence. The fitted relationship was then applied to bulk-tank milk measurements from herds without corresponding individual serological results to estimate their seroprevalence. These estimates, together with the corresponding numbers of lactating cattle, were used to generate seropositive counts through binomial sampling, which were subsequently aggregated to obtain canton-week reference seroprevalence.

    \item \textbf{Parameter sampling.}
    Candidate values of $\beta_0$ and $\Psi_F$ were generated from their prior distributions in the initial ABC-SMC generation.

    \item \textbf{Epidemic simulation.}
    For each proposed parameter combination, one stochastic epidemic realisation was generated from week 36 to week 104, providing the numbers of recovered and total cattle in each canton and week.

    \item \textbf{Generation of simulated seroprevalence.}
    The modelled proportion of recovered cattle was used as the underlying seroprevalence probability. A binomial observation model (see eq. \ref{eq:main_simulated_observation}) was then applied using the corresponding canton-week lactating cattle sample sizes to obtain simulated sampled seroprevalence.

    \item \textbf{Parameter inference.}
    Simulated sampled seroprevalence was compared with the reference seroprevalence for the available canton-week observations during weeks 76--104. The resulting sum of squared differences was used as the ABC distances to accept, reject and weight parameter combinations across successive generations. The weighted particles from the final completed generation were retained for posterior inference.
\end{enumerate}

\section{Results}
\begin{figure}[h]
    \centering
    \includegraphics[
        width=\columnwidth,
        keepaspectratio
    ]{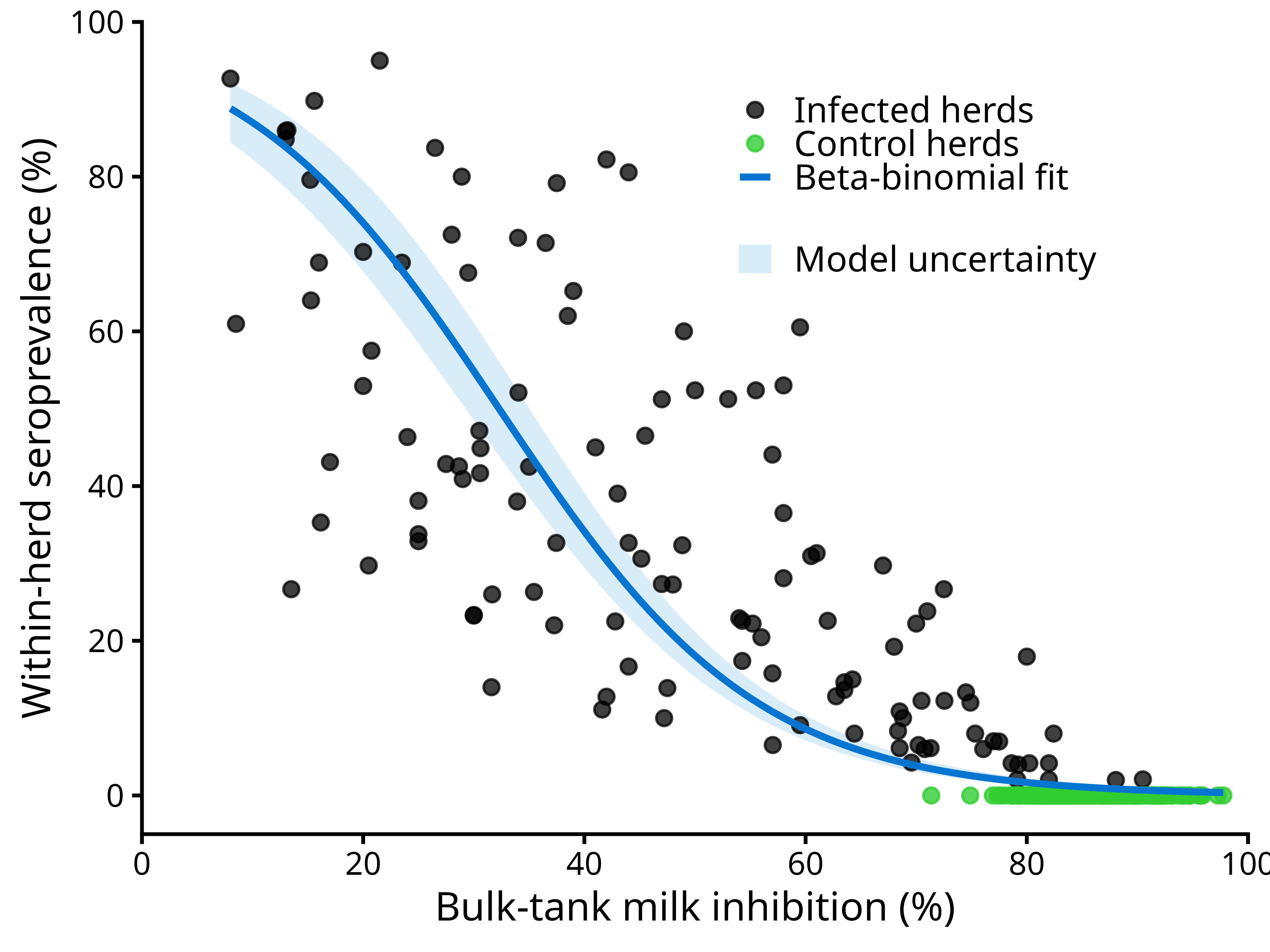}
\caption{\textbf{Relationship between bulk-tank milk inhibition and within-herd seroprevalence, estimated using a beta-binomial regression model.} Points represent observed within-herd seroprevalence and corresponding bulk-tank milk inhibition values for infected herds (black) and control herds (green), for which seroprevalence was assumed to be zero. For control herds with repeated bulk-tank milk measurements, inhibition values were averaged at the herd level. The solid blue line represents the fitted beta-binomial regression relationship, and the light-blue shaded band indicates the associated 95\% confidence interval.}

    \label{fig:estimated_sero}
\end{figure}
\begin{figure*}[h]
    \centering
    \includegraphics[
        width=0.85\textwidth,
        keepaspectratio
    ]{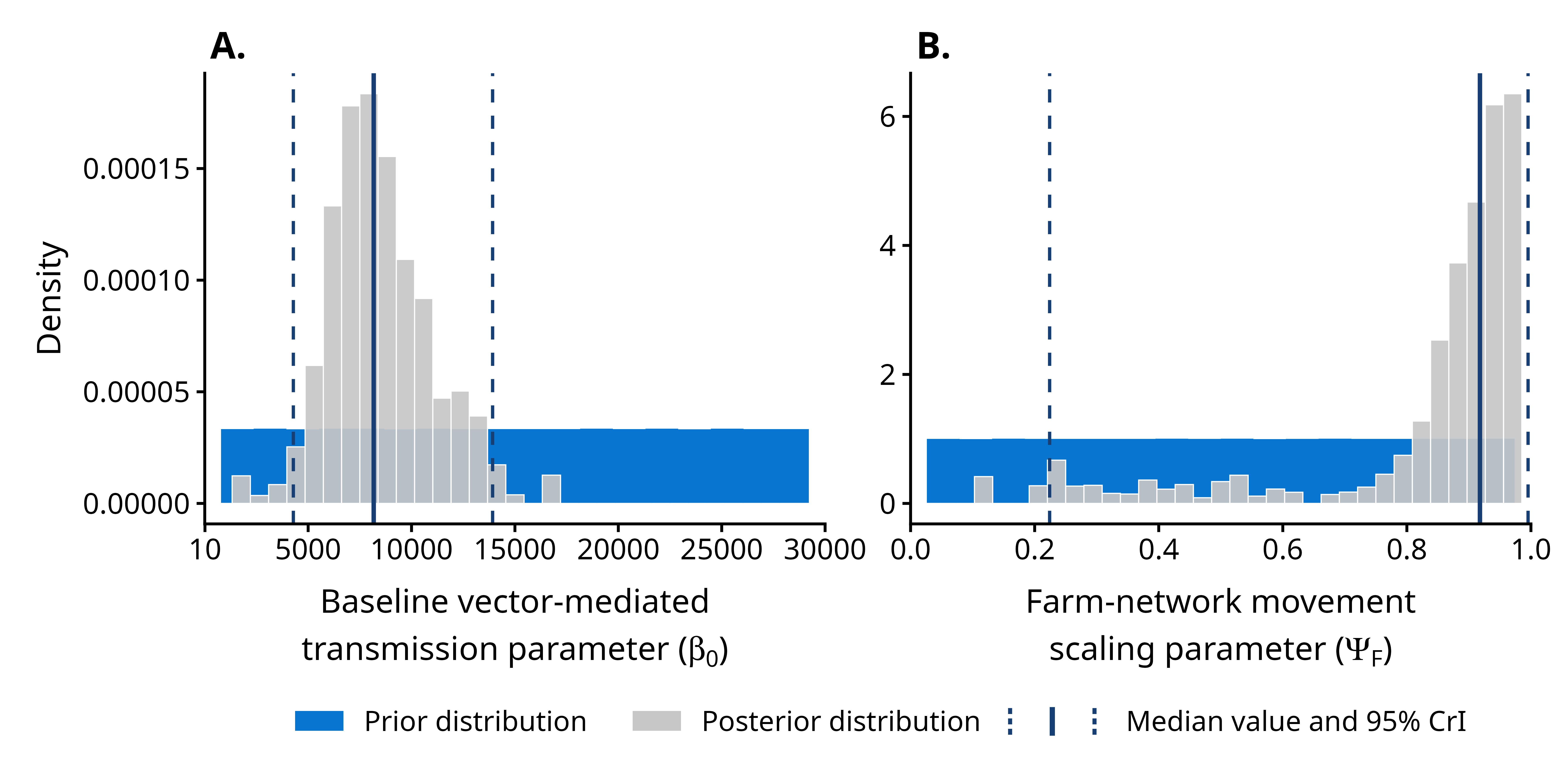}
\caption{\textbf{Comparison of prior and posterior distributions for the two parameters inferred using ABC-SMC.} Panel (A) shows the baseline vector-mediated transmission parameter ($\beta_0$), and panel (B) shows the farm-network movement scaling parameter ($\Psi_F$). Blue histograms represent the prior distributions, while grey histograms represent the posterior distributions obtained from the accepted particles of the final retained ABC-SMC generation. Solid dark-blue vertical lines indicate the weighted posterior medians, and the two dashed dark-blue vertical lines mark the lower and upper bounds of the corresponding 95\% credible intervals (CrI).}
    \label{fig:posterior}
\end{figure*}
\begin{figure*}[h]
    \centering
    \includegraphics[
        width=1.0\textwidth,
        keepaspectratio
    ]{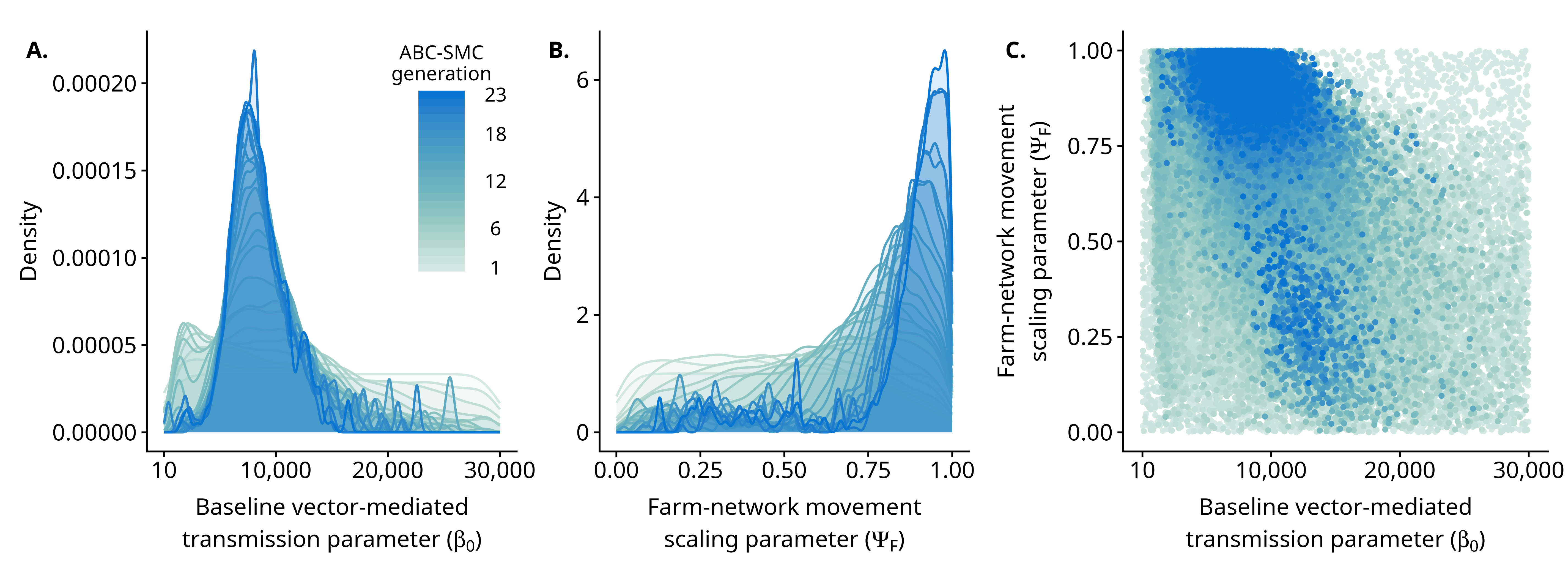}
\caption{\textbf{Evolution of the marginal and joint distributions of the two parameters across successive ABC-SMC generations.} Panels (A) and (B) show the marginal density distributions of the baseline vector-mediated transmission parameter ($\beta_0$) and the farm-network movement scaling parameter ($\Psi_F$), respectively. Panel (C) shows the joint distribution of these parameters, with each point representing an accepted particle. Colours indicate successive ABC-SMC generations, progressing from light teal (earlier generations) to dark blue (later generations), as shown by the colour bar in panel (A). The distributions illustrate the progressive refinement of the posterior approximation as the ABC-SMC acceptance tolerance decreased.}

    \label{fig:pairplot}
\end{figure*}

\subsection{Relationship between bulk-tank milk inhibition and herd seroprevalence}

Bulk-tank milk inhibition was inversely associated with within-herd seroprevalence (Fig. \ref{fig:estimated_sero}). The beta-binomial formulation provided a better fit than the corresponding binomial model, consistent with overdispersion among herds (Supplementary Table~S1). The fitted relationship predicted a within-herd seroprevalence of approximately 74\% at 20\% inhibition, compared with 9\% at 60\% inhibition (Fig. \ref{fig:estimated_sero}). This decrease in predicted seroprevalence with increasing inhibition was reflected in the negative regression coefficient ($-0.0834$ on the logit scale; standard error $=0.0037$, $p<0.001$). Fitted seroprevalence approached zero at high inhibition values. This relationship was subsequently used to estimate herd-level seroprevalence from bulk-tank milk measurements for herds without corresponding individual-level serological data.

\subsection{Posterior estimates of the transmission parameters}

The observed seroprevalence patterns provided clear information on the range of plausible values for the two transmission parameters, although the degree of uncertainty differed markedly between them (Fig. \ref{fig:posterior}). For the baseline vector-mediated transmission parameter, $\beta_0$, the posterior distribution was substantially more concentrated than the prior distribution, with a weighted posterior median of $8{,}171$ and a 95\% credible interval (CrI) of $4{,}284$--$13{,}921$. The narrower posterior distribution suggests that the seroprevalence data helped constrain the baseline vector-mediated transmission parameter, although the width of the 95\% credible interval indicates that uncertainty remained in its estimation (Fig. \ref{fig:posterior}A).

For the farm-network movement scaling parameter, $\Psi_F$, calibration
favoured values towards the upper end of the admissible range (Fig. \ref{fig:posterior}B). The weighted posterior median was $0.918$ (95\% CrI: $0.224$--$0.995$). Within the model, higher values of this parameter correspond to greater cattle exchange between cantons containing pastures belonging to the same farm. The posterior shift therefore suggests that the seroprevalence patterns were consistent with a contribution from cattle movements through the farm network. Nevertheless, the long tail towards lower values and the wide credible interval indicate that considerable uncertainty remained despite the posterior shift towards higher values.

The evolution of the accepted particle distributions across ABC-SMC generations illustrated how calibration progressively refined the parameter combinations retained (Fig. \ref{fig:pairplot}). Later distributions became concentrated around the main peak for $\beta_0$ and near the upper boundary for $\Psi_F$. The joint plot also showed combinations in which lower values of $\Psi_F$ were accompanied by higher values of $\beta_0$. This pattern suggests a possible compensatory relationship during calibration: lower modelled cattle movement could be partly offset by greater vector-mediated transmission. This pattern suggests that different combinations of vector-mediated transmission and farm-network movement could produce similar seroprevalence patterns, although the distributions across generations do not establish parameter dependence in the final posterior.

\subsection{Assessment of parameter recovery and model predictions}
\label{subsec:model_assessment}
The ability of the ABC-SMC procedure to reproduce simulated epidemic patterns varied across the 50 pseudo-observed datasets (Supplementary Fig.~S4). For most datasets, simulations generated from the inferred posterior distributions showed greater median discrepancies from the pseudo-observed seroprevalence trajectories than those generated using the known parameter values. Nevertheless, the discrepancies were relatively small for both approaches in many datasets, while a few showed substantially larger differences. In a small number of cases, posterior predictive simulations achieved comparable or lower discrepancies than simulations using the generating parameters. These results indicate that the inferred parameter distributions could reproduce the simulated epidemic patterns to varying degrees, although predictive performance was not consistently as good as that obtained using the generating parameters.

\begin{figure*}[h]
    \centering
    \includegraphics[
       width=0.8\textwidth,
        keepaspectratio
    ]{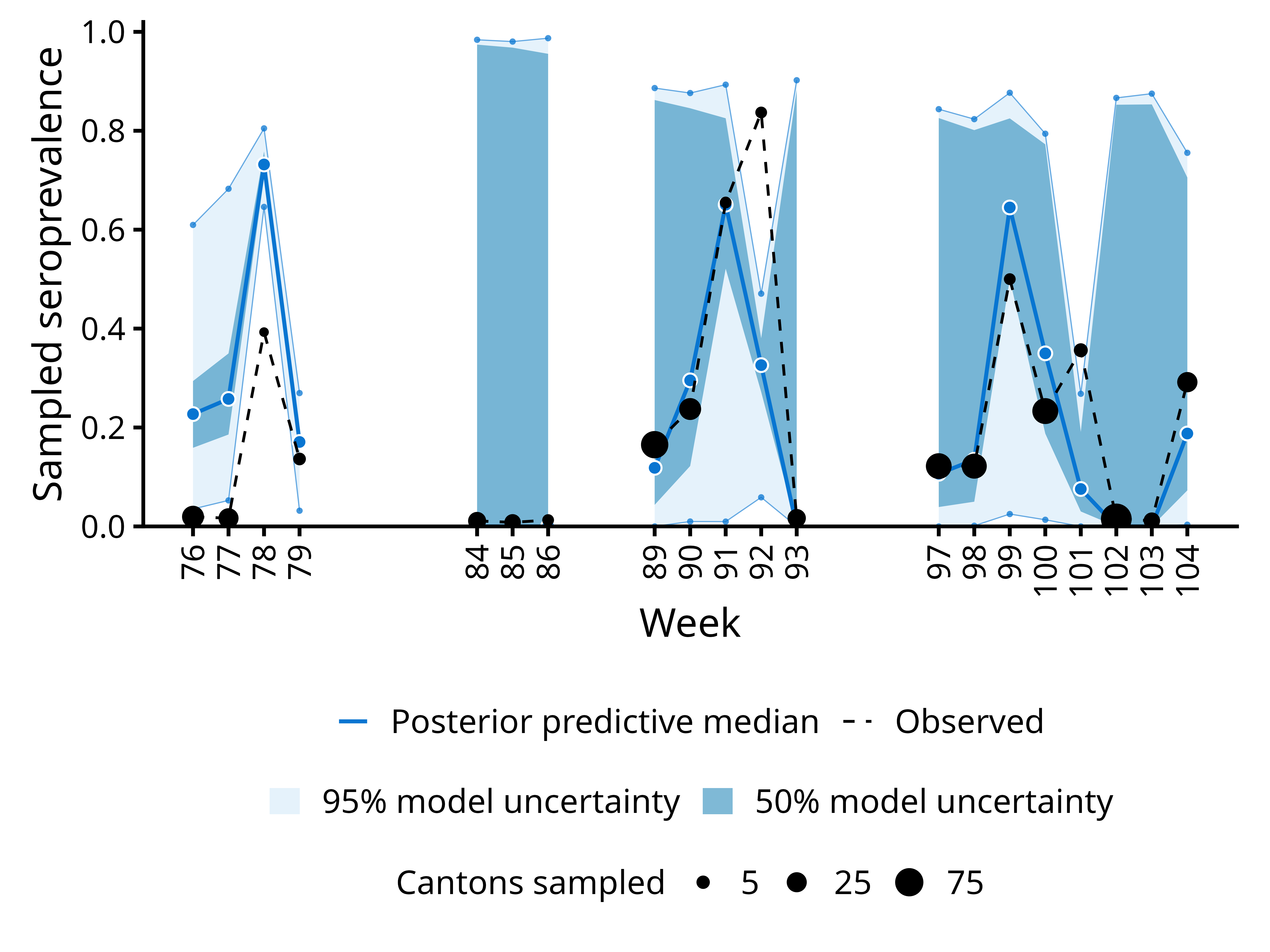}

    \caption{\textbf{Temporal comparison of observed and posterior predictive weekly median seroprevalence across French cantons.} The solid blue line and points represent the posterior median of the simulated trajectory medians across cantons. The dark-blue and light-blue shaded regions indicate the central 50\% and 95\% posterior predictive intervals, respectively. The black dashed line and points represent the observed weekly median seroprevalence across cantons, with point size indicating the number of distinct cantons sampled each week. Larger points correspond to weeks with more sampled cantons. Shaded intervals are displayed only for weeks with available observed data, and gaps indicate weeks for which observed data were unavailable.}

    \label{fig:pretemp}
\end{figure*}
\begin{figure*}[!t]
    \centering
    \includegraphics[
        width=0.95\textwidth,
        height=0.72\textheight,
        keepaspectratio
    ]{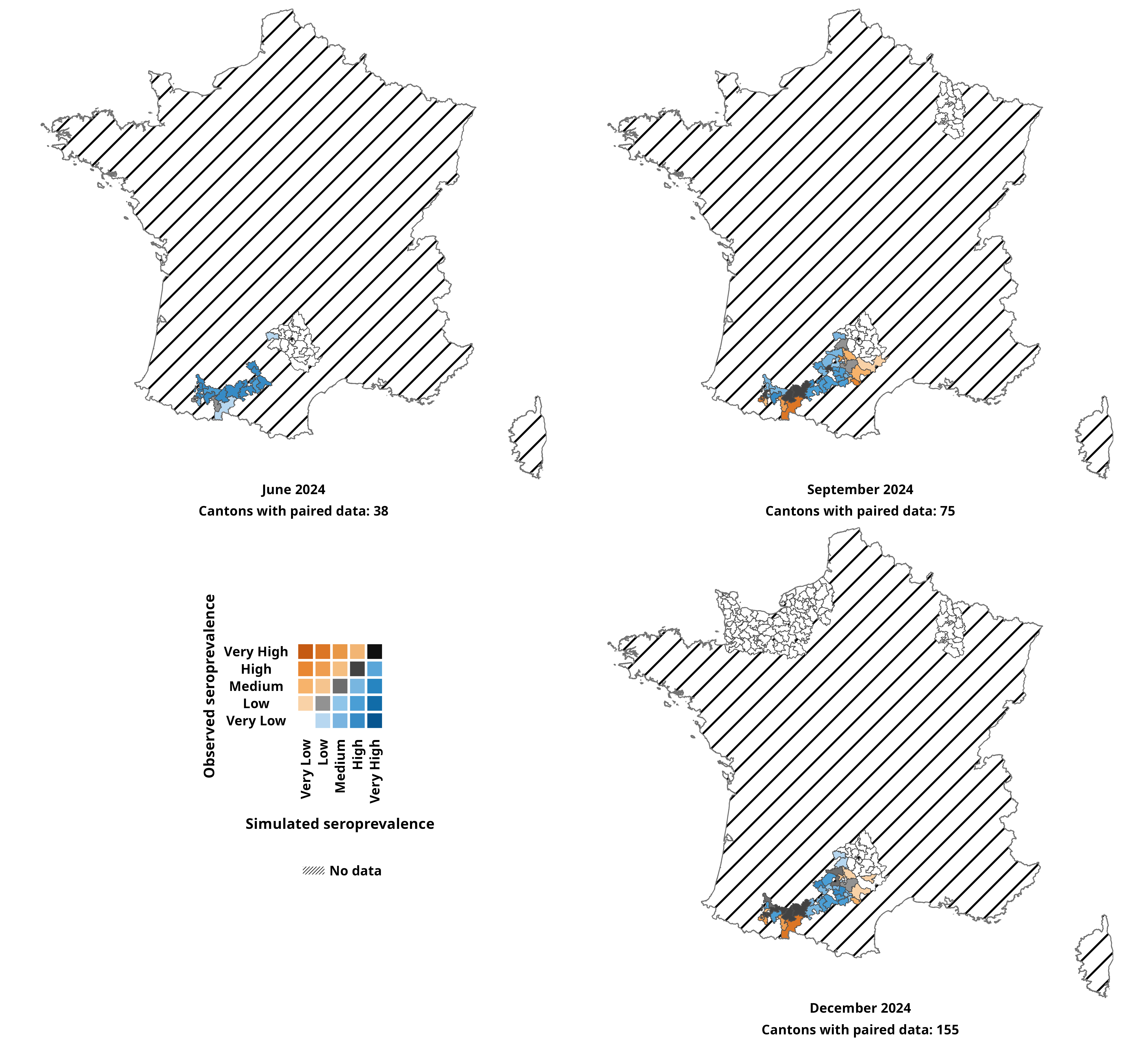}
\caption{\textbf{Spatial comparison of observed and posterior simulated EHDV-8 seroprevalence across French cantons in June, September and December 2024.}
The three monthly bivariate maps compare observed seroprevalence with the corresponding posterior simulated median, classified into five categories (very low, low, medium, high and very high) using fixed 20-percentage-point intervals (0--20\%, $>$20--40\%, $>$40--60\%, $>$60--80\% and $>$80--100\%).
The 25 bivariate colour combinations distinguish higher observed seroprevalence (orange), higher simulated seroprevalence (blue), and agreement between categories (white-to-black diagonal).
Cantons without paired observed and simulated data are indicated by black diagonal hatching on a white background.
The number of cantons with paired data is reported below each monthly map.}
    \label{fig:prespaMain}
\end{figure*}

When applied to the observed EHDV-8 data, posterior predictive simulations reproduced several features of the temporal variation in seroprevalence across sampled cantons (Fig.~\ref{fig:pretemp}). The predicted weekly medians reflected periods of low seroprevalence interspersed with increases broadly consistent with those observed. However, the magnitude of these increases was not always accurately reproduced. In particular, the model overestimated the observed median around week 78 and underestimated it around week 92, with further discrepancies during the later observation weeks. The 90\% posterior predictive intervals were wide at several time points, indicating considerable variation among simulated epidemic trajectories despite calibration to the observed seroprevalence data.

Spatial agreement between observed and simulated seroprevalence
varied across regions and months (Fig.~\ref{fig:prespaMain}).
The use of
fixed seroprevalence categories provided a consistent basis
for identifying where observed and simulated patterns agreed
and where local differences remained. In northern and north-eastern France, several sampled cantons
fell within the lowest seroprevalence category for both
observations and simulations, indicating agreement in areas
with low seroprevalence. In southwestern France, which included
the model's initially infected cantons, simulated seroprevalence
exceeded observations in several cantons in June 2024.
However, these differences narrowed in some cantons over
subsequent months as observed data increased.
By September and December, the spatial patterns were more
heterogeneous, with agreement in some southwestern cantons
and differences in both directions in others. The number of herds sampled also varied between cantons, providing additional context for interpreting these local
differences. Furthermore, the geographical
coverage of paired observations varied between months,
which may influence comparisons of spatial patterns.
The complete monthly comparison, including out-of-sample
predictions for January 2025, is presented in
Supplementary Fig.~S5.

Taken together, the simulation-based assessment and posterior predictive checks showed that the calibrated model was able to reproduce important features of EHDV-8 seroprevalence dynamics, although its performance varied across simulated datasets, sampling times and geographical areas. The model broadly captured temporal changes in seroprevalence and reproduced several regional patterns observed in the serological data. However, discrepancies remained in the magnitude of seroprevalence at particular weeks and in its geographical distribution, alongside substantial variation among posterior predictive simulations. These differences should also be interpreted in light of the changing geographical coverage of the observations, as the same cantons were not necessarily sampled throughout the study period. Overall, the model provided a reconstruction of the broad spatial and temporal patterns of EHDV-8 circulation in France, while the finer-scale heterogeneity of the observed epidemic remained less consistently reproduced.

\section{Discussion}

In this study, we developed and calibrated a stochastic, spatially explicit, multilayer framework to reconstruct EHDV-8 transmission among cattle in mainland France. The framework combined contemporary data on cattle demography and movements, pasture distribution, predicted \textit{Culicoides} abundance, temperature-dependent vector processes, and serological surveillance. Three network layers distinguished short-range vector dispersal, non-commercial cattle movements between pastures belonging to the same farm, and recorded cattle trade. Calibration substantially narrowed the plausible range of the baseline vector-mediated transmission parameter, $\beta_0$, and favoured relatively high farm-network movement intensities, although $\Psi_F$ remained uncertain. Because different epidemic trajectories can arise from the same parameter values in a stochastic model, we also tested the calibration procedure using simulated epidemics generated with true parameters. The resulting posterior distributions reproduced the corresponding epidemic patterns to varying degrees across datasets. When applied to the observed data, the calibrated model reproduced broad temporal and spatial features of the epidemic, including low seroprevalence in several sampled cantons in northern and north-eastern France, while local differences remained across regions and sampling periods.

The original BTV framework provided a biologically appropriate starting point because BTV and EHDV are closely related orbiviruses transmitted by \textit{Culicoides} biting midges \cite{courtejoie2019network}. However, its application to EHDV-8 required substantial reconstruction rather than a simple replacement of disease-specific parameters. We updated the spatial structure to current canton boundaries and incorporated contemporary cattle populations, births, commercial movements, pasture locations, farm structure, temperature, and predicted vector abundance. Unlike the original BTV model, we did not explicitly represent compliance with cattle movement restrictions, including those introduced within 150 km regulated zones during the EHDV-8 epidemic. Instead, we relied on observed cattle movements recorded in the BDNI, which reflected movement patterns under the restrictions in place, although this approach did not allow their specific effects on disease spread to be assessed. A particular strength of the resulting framework was the explicit representation of epidemiologically distinct pathways of between-canton spread. Previous EHDV modelling studies have mainly quantified transmission potential or climate-driven vectorial capacity \cite{gubbins2024using,fairbanks2024modelling}, whereas the present framework represented local vector dispersal, non-commercial cattle movements, and trade within a common dynamic model. This structure allowed the contribution of vector-mediated transmission and farm-network movements to be evaluated against the same spatio-temporal observations. Nevertheless, because the EHDV and original BTV applications differed in host composition, spatial resolution, vector-related parameterisation, surveillance data, and epidemic context, their numerical parameter estimates should not be compared directly.

A second major strength of this study was the use of bulk-tank milk surveillance to complement the limited individual-level serological data. By relating milk inhibition to within-herd seroprevalence, we reconstructed seroprevalence for a substantially larger set of herds and dates and generated the canton-week dataset used for model calibration. This extended the epidemiological information available beyond reported clinical cases, which capture only a subset of infections. However, surveillance coverage remained uneven across space and time. Repeated sampling over additional months and across a broader and more geographically balanced range of herds would strengthen the reconstruction of epidemic trajectories and the validation of spatial model predictions.

The simulation-based assessment clarified how much information these observations provided about the two inferred parameters. For the observed EHDV-8 data, the posterior distribution of $\beta_0$ was substantially more concentrated than its prior, with a median of $8{,}171$ (95\% CrI: 4,284--13,921). By contrast, the posterior for $\Psi_F$ was shifted towards the upper end of its prior range but remained broad, with a median of $0.918$ (95\% CrI: 0.224--0.995). The accepted parameter combinations also suggested that lower farm-network movement could be partly offset by greater vector-mediated transmission. The available seroprevalence information therefore constrained the overall level of vector-mediated transmission more strongly than the intensity of farm-network movements and did not fully separate their respective contributions. Importantly, the posterior concentration of $\Psi_F$ near the upper end of its prior should not be interpreted as a direct estimate of the proportion of cattle moved each week. Rather, it indicates that, within the current model structure, parameter combinations involving relatively high movement intensity through farm-network connections were more consistent with the observed seroprevalence patterns. This result identifies the type of additional information required to distinguish these processes more precisely, particularly direct data on the timing and volume of cattle movements between pastures belonging to the same farm.

Posterior predictive simulations reproduced several changes in the weekly median seroprevalence across sampled cantons, but the magnitude of individual increases was not always captured and uncertainty was wide at several sampling times. Spatially, agreement between observed and simulated patterns varied across regions and over time, but the absence of a consistent tendency towards over- or underestimation suggests that the model captured important geographical features of the epidemic despite local discrepancies. Comparisons among months should additionally be interpreted in light of changes in surveillance coverage, because the number and geographical distribution of sampled cantons varied over time. Overall, the model was most informative about broad spatial and temporal epidemic patterns and now provides a calibrated basis for more targeted analyses of transmission pathways and intervention strategies.

A further strength was the integration of predicted spatio-temporal variation in \textit{Culicoides} abundance with temperature-dependent biting activity, survival, and completion of the extrinsic incubation period \cite{hammami6461683climate}. The non-Markovian formulation retained the delayed contribution of vector cohorts infected during preceding weeks while avoiding an explicit vector population model requiring additional poorly informed parameters. The normalised abundance index represented relative variation in vector availability across cantons and weeks, while pasture cover provided a parsimonious proxy for opportunities for cattle--vector contact. As the abundance predictions were derived from livestock-farm trapping data and environmental covariates, some effects associated with pasture coverage and elevation may already be reflected in the relative-abundance index. Similarly, variation in trap catches may partly reflect vector activity, potentially overlapping with the separate temperature-dependent biting-rate term. Although these additional factors were retained to represent distinct aspects of vector-mediated transmission, the extent of their overlap has not been assessed separately. Short-range vector dispersal was represented through the pasture network, and the epidemic was seeded only in the seven cantons where EHDV-8 was first confirmed, without imposing subsequent exogenous introductions. This provided a stringent test of whether the represented internal transmission processes could generate the subsequent epidemic pattern. Remaining discrepancies, particularly in northern areas, could nevertheless reflect longer-range or windborne vector dispersal, unobserved external introductions, or local vector heterogeneity not captured by the current formulation \cite{burgin2013investigating,aguilar2019possible}. Where suitable data become available, incorporating windborne dispersal estimates, species-specific vector information, or genomic and cross-border surveillance data would allow these alternative mechanisms to be evaluated directly and retained only if they improve model performance.

The later part of the study period overlapped with the state-funded vaccination campaign initiated in September 2024. The herds contributing seroprevalence data to calibration were unvaccinated, limiting direct confounding between vaccine-induced antibodies and natural infection. Early uptake was below the campaign target, with approximately one-third of cattle in the vaccination zone having received a first dose by 17 October 2024, compared with a target of at least 70\% \cite{draaf2024vaccination}. The later spread into northern France overlapped more closely with the vaccination campaign than the initial epidemic in south-western France, although its contribution to the observed epidemic pattern cannot be distinguished from other sources of variation. Future analyses incorporating spatially and temporally resolved vaccination coverage could explicitly account for its population-level effects.

Taken together, this study establishes a calibrated national-scale platform that links serological surveillance, livestock connectivity, environmental suitability, and vector dynamics within a common mechanistic framework. Its immediate next application will be the evaluation and optimisation of vaccination strategies, including campaign timing, geographical targeting, allocation of limited vaccine doses, and associated costs, together with comparison against alternative control measures. Incorporating explicit vaccination coverage and continuing longitudinal serological surveillance will allow these strategies to be evaluated under realistic epidemiological uncertainty and will move the framework towards prospective decision support for future EHDV outbreaks.

\section*{CRediT authorship contribution statement}

\textbf{Viney Kumar:} Conceptualisation, Methodology, Software, 
Formal analysis, Investigation, Data curation, Visualisation, 
Writing -- original draft, Writing -- review \& editing. \textbf{Ga\"el Beaun\'ee:} Conceptualisation, Methodology, Formal analysis, Investigation, Visualisation, Supervision, Writing -- review \& editing. \textbf{Sandie Arnoux:} Formal analysis, conceptualisation, Writing -- review \& editing. \textbf{Fabien Corbi\`ere:} Data resources, Validation, Writing -- review \& editing. \textbf{Beno\^it Durand:} Resources, Methodology, Supervision, 
Formal analysis, Validation, Writing -- review \& editing. \textbf{H\'el\`ene Cecilia:} Conceptualisation, Methodology, Formal analysis, Investigation, Visualisation, Project administration, Supervision, Writing -- review \& editing.

\section*{Funding}
This research was funded by the French Ministry of Agriculture and Food (grant number C-2024-117).

\section*{Declaration of competing interest}

The authors declare that they have no known competing financial interests  or personal relationships that could have appeared to influence the work  reported in this paper.

\section*{Acknowledgement}
We thank Pachka Hammami for kindly providing the Culicoides abundance predictions used in this study and for helpful discussions regarding the data. We are grateful to the INRAE MIGALE bioinformatics facility (MIGALE, INRAE, 2020. Migale bioinformatics Facility, doi: 10.15454/1.5572390655343293E12) for providing help and/or computing and/or storage resources.


\bibliography{example}

\end{document}